# Design and experiment of electronic speckle shearing phase-shifting pattern interferometer*

**XU Tian-hua**\*\*, **JING Chao, JING Wen-cai, ZHANG Hong-xia, JIA Da-gong, and ZHANG Yi-mo**

*College of Precision Instrument & Optoelectronics Engineering, National Education ministry Key Laboratory of Optoelectronics Information and Technical Science, Tianjin University, Tianjin 300072, China*

An electronic speckle shearing phase-shifting pattern interferometer (ESSPPI) based on Michelson interferometer was based in this paper. A rotatable mirror driven by a step motor in one of its reflective arm is used to generate an adjustable shearing and the mirror driven by piezoelectric transducer (PZT) in the other reflective arm was used to realize phase-shifting. In the experiments, the deformation of an aluminum plate with the same extern-force on different positions and different forces on the same position is measured. Meanwhile, the phase distribution and phase-unwrap image of the aluminum plate with the extern-force on its center position is obtained by the four-step phase-shifting method.

The electronic speckle shearing phase-shifting pattern interferometer (ESSPPI)[1] is a precise measuring technology developed on electronic speckle interferometry[2,3], it has the merits of high accuracy, non-destructive testing and full-field measurement. Compared with other testing method, ESSPPI also has the advantages of vibration-immunity[4,5], sensitivity-adaptable[6] and direct measurements of the displacement[7,8]. The electronic speckle interferometric method with phase-shift technique[9] was firstly proposed by Johansson in 1989. In this paper, a new electronic speckle shearing phase-shifting pattern interferometer based on Michelson interferometer is designed, the shearing interferometry and phase-shift technique are integrated in the electronic speckle measurement.

The scheme of the electronic speckle shearing phase-shift pattern interferometer system is shown in Fig.1. It consists of four parts: light source, Michelson interferometer, central controlling unit, image acquisition module. The main structure is the Michelson interferometer with a rotatable arm driven by stepping motor and a mobile arm driven by PZT.

The working principle of the testing system is as follows: The continuous light of 532 nm from LD is collimated onto the surface of the interested object, which is deformed by the extern force. The coherent light is reflected into the Michelson interferometer by the object, and divided into two beams which transmits in different interferometric arms then. The speckle interferogram is received by the CCD. In order to generate the shearing, the mirror in one of reflective arm needs to rotate by a little angle, which enables two adjacent points image on the same point in the interferogram. The phase shifting is realized by the movement of the mirror driven by PZT in the other reflective arm.

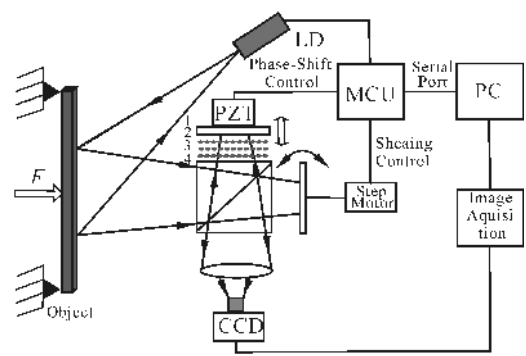

MCU:Micro controller unit,LD:Laser Diode

**Fig.1 The diagram of electronic speckle shearing phase-shifting pattern interferometry**

The optical power distribution in the CCD, before object is deformed, can be described as the following expression[10,11]:

$$I_1 = 2|A|^2 (1 + \cos \phi) \qquad (1)$$

where $\phi$ is the phase difference between two arbitrary adjacent points on the surface of the object, and it obeys the sto-chastic distribution.

\*   This work has been supported by the National Natural Science Fund of China( No. 60577013), and program for New Century Excellent Talents in University, Ministry of Education, China

\*\*   E-mail: xutianhua0629@yahoo.com.cn

After the object is deformed, the optical power distribution can be represented as:

$$I_2 = 2|A|^2 \left[1 + \cos\left(\phi + \Delta\phi\right)\right] \tag{2}$$

where $\Delta\phi$ is the phase difference generated by shearing pattern, and it is related with the derivative of deformation:

$$\Delta\phi = \frac{4\pi}{\lambda} \cdot \frac{\partial w}{\partial x} \, dx \tag{3}$$

where $\dfrac{\partial w}{\partial x}$ is the derivative of the deformation, dx is the value of the shearing.

The speckle shearing pattern interferogram can be obtained by equation (1) and equation (2):

$$I = I_2 - I_1 = 4|A|^2 \sin\left(\phi + \frac{\Delta\phi}{2}\right) \sin\left(\frac{\Delta\phi}{2}\right) \tag{4}$$

In equation (4), $\phi$ is a random phase, therefore $\sin(\phi + \Delta\phi/2)$ is a random value. The optical power distribution changes according to $\sin(\Delta\phi/2)$, thus the speckle shearing interferogram indicates the deformation of the object.

In the experiment, the interested object is a rectangular aluminum plate with a size of 220 mm × 165 mm. First of all, series of electronic speckle shearing interferograms of deformation on the aluminum plate are obtained, when different extern forces are imposed on the center position of the plate. With increasing forces, the corresponding interferograms are recorded in Fig.2 from (a) to (f). According to these figures, it is found that the fringes of the speckle interferogram become denser with the lager forces, which causes more deformation.

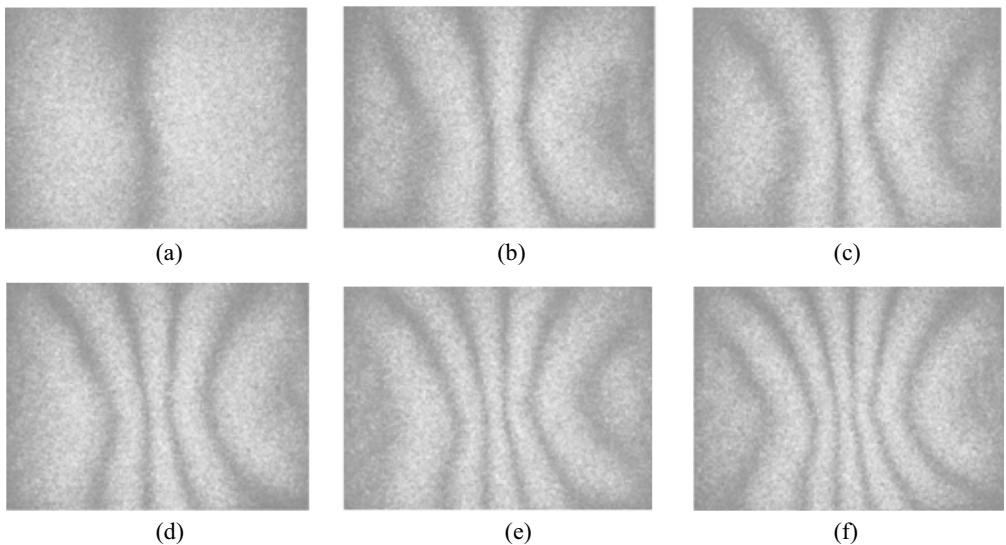

(a)          (b)          (c)

(d)          (e)          (f)

**Fig.2  Speckle interferogram of deformation on aluminum plate with different forces**

When the same extern force is imposed on different positions of the aluminum plate, the speckle shearing interferograms are shown in the Fig.3. The speckle pattern fringes

The phase distribution and phase-unwrap image are obtained with the four-step phase-shift technology, when the concentrate on the position of force exerting, which is also the position of the maximum deformation on the plate.

extern-force inflicts on the center position of the aluminum plate. The experimental results are shown in Fig.4 and Fig.5.

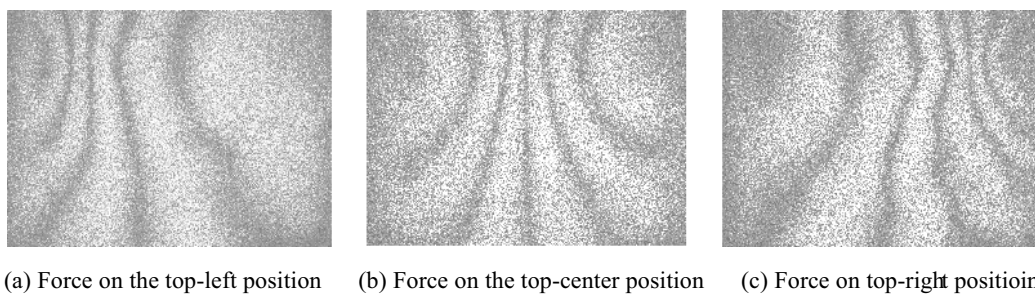

(a) Force on the top-left position     (b) Force on the top-center position     (c) Force on top-right positioin

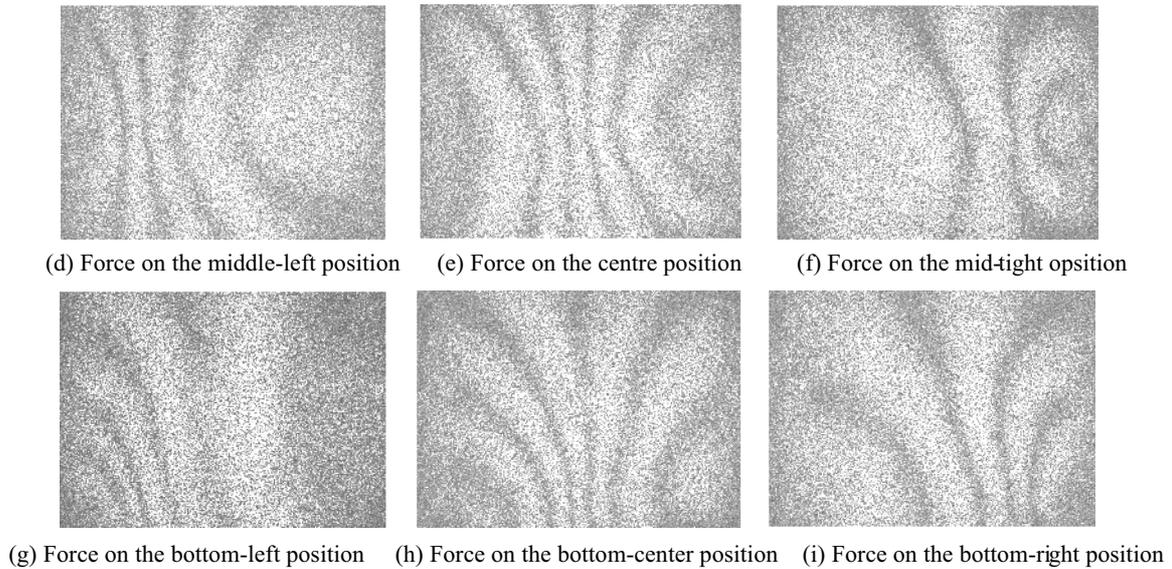

(d) Force on the middle-left position          (e) Force on the centre position          (f) Force on the mid-right opsition

(g) Force on the bottom-left position          (h) Force on the bottom-center position          (i) Force on the bottom-right position

**Fig.3  Speckle interferogram of deformation on aluminum plate with force on different positions**

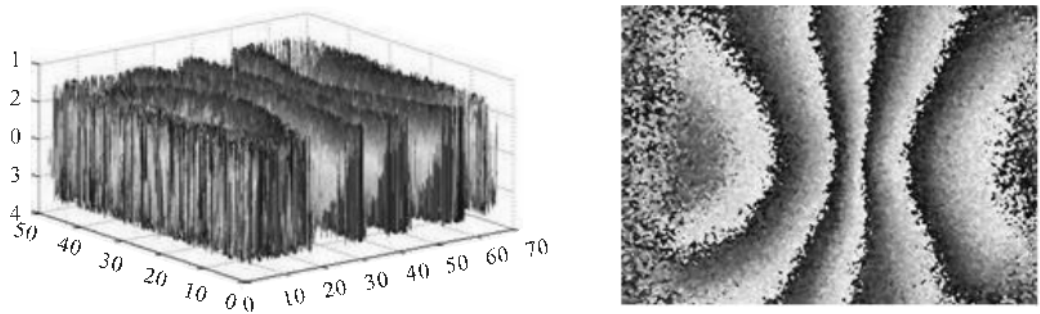

**Fig.4  Experimental result of phase distribution on aluminum plate**

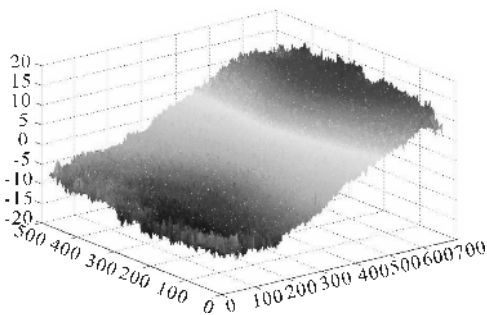

**Fig.5  Experimental result of phase unwrap on aluminum plate**

In this paper, with the integration of the shearing interferometry and phase-shift technique, a new electronic speckle shearing phase-shifting pattern interferometer based on Michelson interferometer is designed. Deformation of an aluminum plate with the same extern-force on different positions and different extern-forces on the same position is measured, and the phase distribution and its unwrapped image of the aluminum plate with the extern-force on its center position is obtained.